\documentclass[a4paper]{article}
\usepackage{graphicx} 
\usepackage[english]{babel}
\usepackage[latin1]{inputenc} 
\usepackage[bf,small,nooneline]{caption}
\usepackage[round]{natbib}

\newcommand{\dfrac}[2]{\frac{\displaystyle#1}{\displaystyle#2}}
\newcommand{\pder}[2]{\frac{\displaystyle\partial#1}{\displaystyle\partial#2}}

\title{An Energy Conserving Parallel Hybrid Plasma Solver}
\date{December 2, 2010}
\author{M.\ Holmstr{\"o}m\thanks{Swedish Institute of Space Physics, PO~Box~812, SE-98128~Kiruna, Sweden. (\texttt{matsh@irf.se})}}

\begin{document}

\maketitle

\begin{abstract}
We investigate the performance of a hybrid plasma solver on 
the test problem of an ion beam.  The parallel solver is based on cell 
centered finite differences in space, and a predictor--corrector 
leapfrog scheme in time.  
The implementation is done in the FLASH software framework. 
It is shown that the solver conserves energy well over time, 
and that the parallelization is efficient (it exhibits weak scaling). 
\end{abstract}

\section{Introduction}
Modeling of collisionless plasmas are often done using fluid 
magnetohydrodynamic models (MHD).  The MHD fluid approximation is however 
questionable when the gyro radius of the ions are large compared to the 
spatial region that is studied.  On the other hand, kinetic models that 
discretize the full velocity space, or full particle in cell (PIC) models 
that treat ions and electrons as particles, are very computational expensive. 
For problems where the ion time- and spatial scales are of interest, 
hybrid models provide a compromise.  In such models, the ions are 
treated as discrete particles, while the electrons are treated as a 
(often massless) fluid.  This mean that the electron time- and spatial 
scales do not need to be resolved, and enables applications such as 
modeling of the solar wind interaction with planets. 
For a detailed discussion of different models, see \citet{led08}.

Here we present an finite difference implementation of a hybrid model 
in the FLASH parallel computational framework, along with test cases 
that show that the implementation scales well and conserve energy well. 

\section{The hybrid equations}
In the hybrid approximation, ions are treated as particles, 
and electrons as a massless fluid. 
In what follows we use SI units. 
The trajectory of an ion, $\mathbf{r}(t)$ and $\mathbf{v}(t)$, 
with charge $q$ and mass $m$, is computed from the Lorentz force, 
\[
  \dfrac{d\mathbf{r}}{dt} = \mathbf{v}, \quad
  \dfrac{d\mathbf{v}}{dt} = \dfrac{q}{m} \left( 
    \mathbf{E}+\mathbf{v}\times\mathbf{B} \right), 
\]
where $\mathbf{E}=\mathbf{E}(\mathbf{r},t)$ is the electric field, 
and $\mathbf{B}=\mathbf{B}(\mathbf{r},t)$ 
is the magnetic field.  The electric field is given by 
\[
  \mathbf{E} = \dfrac{1}{\rho_I} \left( -\mathbf{J}_I\times\mathbf{B} 
  +\mu_0^{-1}\left(\nabla\times\mathbf{B}\right) \times \mathbf{B} 
  \right) - \nabla p_e, 
\]
where $\rho_I$ is the ion charge density, 
$\mathbf{J}_I$ is the ion current, 
$p_e$ is the electron pressure, and
$\mu_0=4\pi\cdot10^{-7}$ is the magnetic constant. Then 
Faraday's law is used to advance the magnetic field in time, 
        \[
          \pder{\mathbf{B}}{t} = -\nabla\times\mathbf{E}. 
        \]

\section{A cell centered finite difference hybrid solver}
We use a cell-centered representation of the magnetic field on a uniform 
grid.  All spatial derivatives are discretized using standard second order 
stencils.  Time advancement is done by a predictor-corrector leapfrog method 
with subcycling of the field update, denoted cyclic leapfrog (CL) 
by \citet{mat94}. An advantage of the discretization is that the 
divergence of the magnetic field is zero, down to round off errors. 
The ion macroparticles (each representing a large number of real particles) are deposited 
on the grid by a cloud-in-cell method (linear weighting), and 
interpolation of the fields to the particle positions are done by 
the corresponding linear interpolation. 
Initial particle positions are drawn from a uniform distribution, 
and initial particle velocities from a Maxwellian distribution. 
Further details of the algorithm can be found in \citet{Enumath09}. 

\section{Implementing the solver in FLASH}
We use an existing software framework, FLASH, developed at 
the University of Chicago \citep{Fryxell00}, that implements a block-structured 
adaptive (or uniform) Cartesian grid and is parallelized using 
the Message-Passing Interface (MPI) library for communication. 
It is written in Fortran 90, well structured into modules, and open source. 
Output is handled by the HDF5 library, providing parallel I/O. 
Although the FLASH framework has mostly been used for fluid 
modeling, it has support for particles which we have used to 
implement a hybrid solver using the latest version of the framework, 
FLASH3. 

The advantage of using an existing framework when implementing a solver 
is that all grid operations, parallelization and file handling is done 
by standard software calls that have been well-tested. 
Also, there is an existing infrastructure for parameter files and 
setup directories that simplifies code handling.  The concept of a 
setup directory is that one can place modified versions of any 
routine in the directory, and this new version will be used during 
the build process.  This is an easy way to handle different versions 
of code for different runs of the solver. 

In particular, many of the basic operations needed for a PIC code are 
provided as standard operations in FLASH: 
\begin{itemize}
\item Deposit charges onto the grid: 
      \texttt{call Grid\_mapParticlesToMesh()}
\item Interpolate fields to particle positions: 
      \texttt{call Grid\_mapMeshToParticles()}
\item Ghost cell update for all blocks: \texttt{call Grid\_fillGuardCells()}
\end{itemize}

The advantage of a parallel solver is the ability to handle larger 
computational problems than on a single processor, both in terms 
of computational time and memory requirements. 
This is especially important for PIC solvers which are computational 
intensive compared to fluid solvers.  Since we typically have 
10--100 particles per cell, the computational work will be dominated 
by operations on the particles: moving the particles and grid-particle 
operations. 

\begin{figure}[ht]
%% Grayscale version of SNIC figure. 
%%  shared/text/papers/snic10/f3scaling-snic.odg
\begin{center}
  \includegraphics[width=0.6\columnwidth]{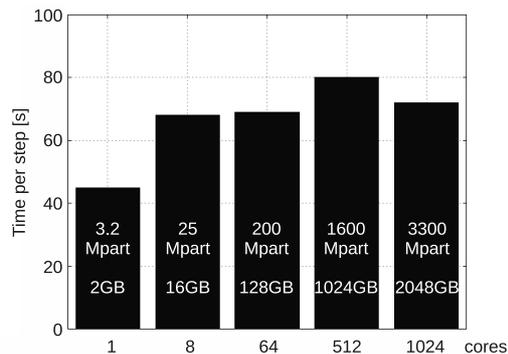}
\end{center}
\caption{Timings for the hybrid solver, illustrating that the code scales well for large problems. We see that the wall clock time required for a time step for this particular test is roughly constant around 70 seconds independent of problem size.  Total number of particles, and total available memory is shown on each bar.  All runs were done on the Akka cluster at the High Performance Computing Center North (HPC2N) at Ume\aa\ University, Sweden.  A 672 node cluster with an Infiniband interconnect, where each cluster node has two quad core Intel Xeon processors and 16 GB of RAM.
} \label{fig:scaling} 
\end{figure}

That a code works well in parallel is usually investigated by looking 
at how the code scales.  For the case of strong scaling, a fixed size 
problem is run on different numbers of processors.  Ideally the execution 
time should decrease proportional to the number of processors (linear scaling). 
This is however difficult to achieve in real world applications. 
Sequential parts of the program quickly dominate the execution time. 
An alternative concept is that of weak scaling.  Here the problem size 
is increased at the same time as we add more processors. 
That this implementation of a hybrid solver exhibits weak scaling is 
shown in Fig.~\ref{fig:scaling}. 
For this application, weak scaling is adequate, since we aim to solve 
larger problems using a constant wall clock time. 

\section{Accuracy of the solver}
It is not easy to check the correctness and accuracy of a hybrid solver. 
There are not many non-trivial test problems that have analytical solutions to 
compare with.  One can of course check the results against published 
results but that only gives a crude sanity check of the code. 
One option is to investigate the conservation of total energy. 
This is a crucial quantity since the behavior of a plasma is a 
constant exchange of energy between the kinetic energy of the particles 
and the energy stored in the electromagnetic fields. 
It is also a quantity that easily can be measured for \emph{any} 
simulation, especially if the boundary conditions are periodic. 
We therefore choose to perform numerical experiments with an ion beam, 
a test problem that strongly exhibits this exchange of energy between particles 
and fields, to study the conservation of total energy. 

\subsection{One-dimensional ion beam}
% Winske and Leroy, JGR, 1984.   1D
% Winske, SSR, 1985.             1D
% Matthews, JCP, 1994.           1D, 2D
% Winske and Quest, JGR, 1986.   1D, 2D
A classic plasma model problem is that of an ion beam into a 
plasma~\citep{win84,win85,win86}. 
\citet{mat94} describes a two-dimensional simulation of a low density ion beam 
through a background plasma.  The initial condition has a uniform magnetic 
field, with a beam of ions, number density $n_{b}$, propagating 
along the field with velocity $v_b=10v_A$, where $v_A=B/\sqrt{\mu_0nm_i}$ 
is the Alfv\'{e}n velocity, through a background (core) plasma of 
number density $n_{c}$. 
Both the background and beam ions have thermal velocities $v_{th}=v_A$.  
Here we study what is denoted a \emph{resonant beam} 
with $n_{b}=0.015\,n_c$ and $v_c=-0.2\,v_A$. 
% and a \emph{non-resonant beam} with $n_{b}=0.1\,n_c$ and $v_c=-1.1\,v_A$. 
Electron temperature is assumed to be zero, so the electron 
pressure term in the electric field equation of state is zero. 
The weight of the macroparticles are chosen such that there is 
an equal number of core and beam macroparticles, each beam macroparticles 
thus represent fewer real ions than the core macroparticles. 
The number of magnetic field update subcycles is nine. 

\begin{figure}[ht]
\begin{center}
  \includegraphics[width=0.7\columnwidth]{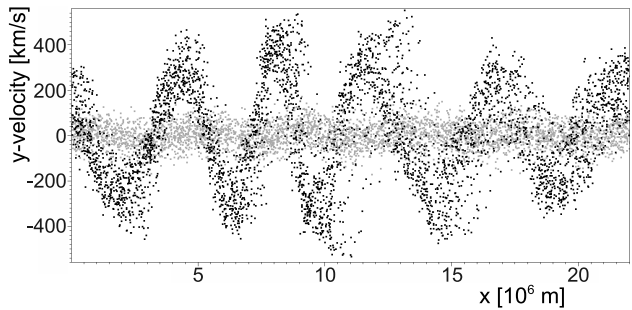}
\end{center}
\caption{Velocity space plot for a one-dimensional ion beam. 
Velocity along the $y$-axis as a function of position at time 
$t=34.6$~s $\approx 20\,\Omega_i^{-1}$.
Each gray dot is a core macro-ion, and each black dot is a beam macro-ion. 
} \label{fig:beam} % hpc2n/pub/v04-part/
\end{figure}

The spatial extent of the domain is 22016~km along the $x$-axis, 
divided up in 256~cells with periodic boundary conditions.  
The core number density is $n_c=7$~cm$^{-3}$. 
The magnetic field magnitude is $B=6$~nT directed along the $x$-axis, 
which give an Alfv\'{e}n velocity of 50~km/s and an ion inertial length 
of $\delta_i$=86~km, where $\delta_i=v_A/\Omega_i$ 
and $\Omega_i=q_iB/m_i$ is the ion gyrofrequency. 
The time step is 0.0865~s = $0.05\,\Omega_i^{-1}$, and the cell 
size is $\delta_i$. The number of particles per cell is 32. 
%% Note that the magnetic field energy plots are different 
%% in Winske SSR 1985 Fig 1. Wrong time scale? 
%% Use time and size of second magnetic energy minimum as target?
%% Or just energy conservation?
In Fig.~\ref{fig:beam} we show a velocity space plot of the 
macro-ions at time $t=34.6$~s $\approx 20\,\Omega_i^{-1}$. 
This can be compared to Fig.~5 in \citet{win84}. 

\begin{figure}[ht]
\begin{center}
  \includegraphics[width=0.7\columnwidth]{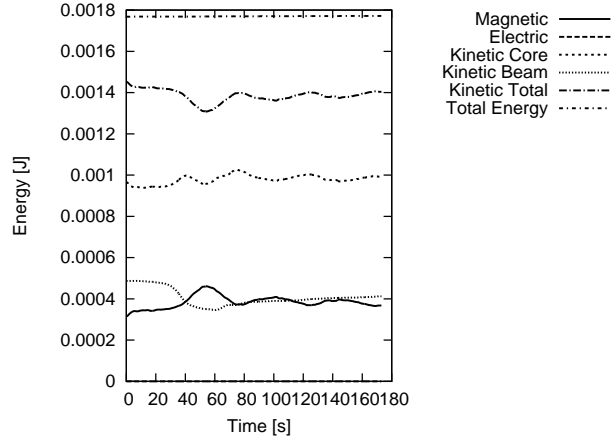}
\end{center}
\caption{Energy partition for a one-dimensional ion beam as a function of time. 
} \label{fig:1dE} % hpc2n/pub/v04-part/astronum-v04-part.gnuplot
\end{figure}

The kinetic energy of the ions, the energy stored in the electromagnetic 
fields, and total energy, as a function of time, 
are shown in Fig.~\ref{fig:1dE}. 
The magnetic and electric field energies are computed as 
\[
   \frac{1}{2\mu_0}\int\left|\mathbf{B}\right|^2 dV, \quad \mbox{ and } \quad 
   \frac{\epsilon_0}{2}\int\left|\mathbf{E}\right|^2 dV, 
\]
respectively. 
Here $\epsilon_0$ is the vacuum permittivity, $\epsilon_0\mu_0c^2=1$. 
Note that the electric field energy is too small to be visible in the figure 
(orders of magnitude smaller than the magnetic field energy). 
%% matsh@vega:/srv/run/hpc2n/pub/v04-part$ gnuplot astronum-v04-part.gnuplot 
%% Keep bounding box: 
%%   $ ps2pdf -dEPSCrop astronum-v04-part.ps astronum-v04-part.pdf
%The relative energy gain at $t=173\approx 100\,\Omega_i^{-1}$ 
%is less than 0.21\%. 
At $t=86$~s $\approx 50\,\Omega_i^{-1}$ the relative energy error is 
less than 0.1\%. 
This can be compared to the error of 11\% given by \citet{mat94}. 
%% matsh@vega:/srv/run/hpc2n/pub/v04-part$ 
%%     /home/matsh/shared/local/FLASH2.5/setups/hybrid/energy.py <flash.dat
%%  Nr of meta particles =  8197.0 
%%          t          Etot           dErel
%%        0.173      0.0017681   -3.62134e-05
%%       86.327     0.00176977    0.000908433
%%      171.962     0.00177181     0.00206027
%%          173     0.00177181     0.00205945

\subsection{Two-dimensional ion beam}
\begin{figure}[ht]
\begin{center}
  \includegraphics[width=0.6\columnwidth]{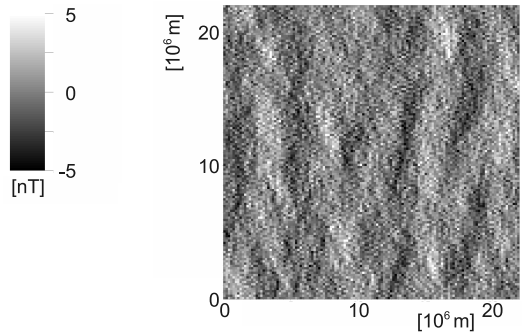}
\end{center}
\caption{The magnetic field $y$-component for a 
two-dimensional ion beam at time $t=77.85$~s $\approx 46\,\Omega_i^{-1}$. 
} \label{fig:2d} % v09b
\end{figure}

\begin{figure}[ht]
\begin{center}
  \includegraphics[width=0.7\columnwidth]{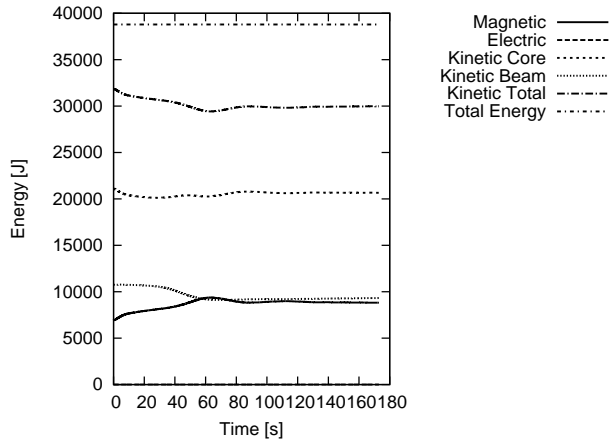}
\end{center}
\caption{Energy partition for a two-dimensional ion beam as a function of time. 
} \label{fig:2dE} % v09b
\end{figure}

In the two-dimensional case we have a square grid with sides of length 
22016~km, and 128 cells in each direction with periodic boundary conditions. 
The time step is 0.0216~s = $0.05\,\Omega_i^{-1}$, and the cell 
widths are $2\delta_i$.  
The number of particles per cell is 16. 
Otherwise the setup is identical to the one-dimensional case. 
% Winske -86 has 262 144 ions = 16 ppc
In Fig.~\ref{fig:2d} we show the magnitude of the magnetic field 
$y$-component at time $t=77.85$~s $\approx 46\,\Omega_i^{-1}$.
This can be compared to Fig.~5 in \citet{win86}. 
The kinetic energy of the ions, the energy stored in the electromagnetic 
fields, and total energy, as a function of time, 
are shown in Fig.~\ref{fig:2dE}. 
%% matsh@vega:/srv/run/hpc2n/v09b$ gnuplot astronum-v09b.gnuplot
%% Keep bounding box: $ ps2pdf -dEPSCrop astronum-v09b.ps astronum-v09b.pdf
%% matsh@vega:/srv/run/hpc2n/v09b$ /home/matsh/shared/local/FLASH2.5/setups/hybrid/energy.py <flash.dat 
%% Nr of meta particles =  262103.0 
%%         t          Etot           dErel
%%     0.04325        38796.5   -1.13894e-06
%%     86.4568        38797.1    1.54511e-05
%%     171.962        38797.7    3.10894e-05
%%         173        38797.7    3.12525e-05
%% Note: Better than v09/ that has more particles: 
%%  Nr of meta particles =  2096567.0 
%%          t          Etot           dErel
%%       0.0865          38758   -4.91439e-07
%%      86.4135        38763.3    0.000134903
%%      171.962        38767.1    0.000233857
%%          173        38767.2    0.000235383
We can note that the fluctuations in magnetic and kinetic energy 
is smaller than in the one-dimensional case, consistent with the 
observations by \citet{win86}. 
The total relative energy gain at $t=173$~s $\approx 100\,\Omega_i^{-1}$ 
is less than 0.004\%. 
This can be compared to \citet{win86}, where the stated energy gain was 
less than 2\% and smoothing of the fields was needed for numerical stability. 

\section{Summary and conclusions}
We have investigated the performance of a hybrid solver on the classical 
test case of an ion beam, in one and two dimensions. 
It is shown that the cell centered finite difference solver 
conserves energy very well.  
The change in total energy over time is more than an order of magnitude 
smaller than for two previously published solvers.  
Also, the solver is numerically stable over long times without any smoothing 
of the fields. 
It is planned that this hybrid solver will be part of a future 
release of the FLASH code. 

\subsection*{Acknowledgments} 
This research was conducted using resources provided by the Swedish National 
Infrastructure for Computing (SNIC) at the High Performance Computing Center 
North (HPC2N), Ume\aa\ University, Sweden.
The software used in this work was in part developed by the 
DOE-supported ASC / Alliance Center for Astrophysical 
Thermonuclear Flashes at the University of Chicago.

\end{document}